\documentclass[a4paper,manyauthors,nocleardouble,COMPASS]{cernphprep}

\RequirePackage[T1]{fontenc}

\RequirePackage{graphicx}
\RequirePackage{newtxtext}
\RequirePackage{newtxmath}
\RequirePackage{flushend}
\RequirePackage[numbers,sort&compress]{natbib}

\usepackage{etex}
\usepackage{amsmath,amssymb,epsfig,fleqn,url,color,here,graphics,graphicx,rotating,multirow,wrapfig}
\usepackage{pbox}
\usepackage{dcolumn}
\usepackage{bigstrut}
\usepackage{soul}
\usepackage{lineno, xcolor}
\usepackage{balance}
\usepackage{lastpage}

\usepackage[section]{placeins}
\usepackage{orcidlink}
\usepackage{tikz}
\usepackage{hyperref}
\hypersetup{
    colorlinks,
    citecolor=blue,
    filecolor=blue,
    linkcolor=blue,
    urlcolor=blue
}
\def\ptv{\textbf{\textit{P}}_{\rm{T}}}
\def\stv{\textbf{\textit{S}}_{\rm{t}}}

\def\gt{g_{\rm{T}}}

\def\Pt{P_{\rm t}}

\def\gevv{\rm{GeV}}
\def\gevcc{\rm{GeV}/\textit{c}}
\def\gevc2{\rm{GeV}/\textit{c}^2}
\def\gev2{(\rm{GeV}/c)^{2}}
\def\pht{\textit{P}_{\rm{T}}}

\def\Acoll{A_{\rm{Coll}}^{h}}
\def\Asiv{A_{\rm{Siv}}^{h}}
\def\kt{\textbf{\textit{k}}_{\rm T}}
\def\ktt{\textit{k}_{\rm T}^{2}}

\def\phiColl{\phi_{\rm{Coll}}}
\def\phiSiv{\phi_{\rm{Siv}}}

\def\Coll{{\rm{Coll}}}
\def\Siv{{\rm{Siv}}}
\def\fuq{f_{1}^{q}}
\def\guq{g_{1}^{q}}
\def\huq{h_{1}^{q}}
\def\Hhuq{H_{1q}^{h}}
\def\Dhuq{D_{1q}^{h}}
\newcommand{\fSiv}[1]{f_{1\rm{T}}^{\perp #1}}
\newcommand{\mfSiv}[1]{f_{1\rm{T}}^{\perp (1) #1}}
\def\Dnn{D_{\rm{NN}}}

\makeatletter
\g@addto@macro\bfseries{\boldmath}
\makeatother
\begin{document}
\begin{titlepage}
\PHnumber{2023-XXX}
\PHdate{\today}

\title{High-statistics measurement of Collins and Sivers asymmetries for transversely polarised deuterons}

\begin{abstract}
New results are presented on a high-statistics measurement of Collins and Sivers asymmetries of charged hadrons produced in deep inelastic scattering of muons on a transversely polarised $^6$LiD target.
The data were taken in 2022 with the COMPASS spectrometer using the 160 \gevv\ muon beam at CERN, balancing the existing data
on transversely polarised proton targets.
The first results from about two-thirds of the new data have total
uncertainties smaller by up to a factor of three compared to
the previous deuteron measurements.
Using all the COMPASS proton and deuteron results, both the transversity and the Sivers distribution functions of the $u$ and $d$ quark, as well as the tensor charge
in the measured $x$-range are extracted. In particular, the accuracy of the $d$ quark results is significantly improved.
\end{abstract}

\Collaboration{The COMPASS Collaboration}
\ShortAuthor{The COMPASS Collaboration}
\vfill
\Submitted{(to be submitted to Phys. Rev. Letters)}
\end{titlepage}

\date{\today}
\maketitle

{
\pagestyle{empty}
\clearpage
}
\clearpage

\setcounter{page}{1}

%
%

\label{sec:intro}

Transverse spin effects observed several decades ago in high-energy hadron interactions~\cite{Bunce:1976yb} originally constituted a challenge to Quantum Chromodynamics (QCD)~\cite{Kane:1978nd}.
Intense theoretical work~\cite{Ralston:1979ys,Artru:1989zv,Jaffe:1991kp} followed, and pioneering results were produced by the HERMES and COMPASS collaborations
in semi-inclusive hadron production measurements of deep inelastic
lepton--nucleon scattering (SIDIS) off transversely polarised proton and
deuteron targets.
The current theoretical description~\cite{Bacchetta:2006tn} involves both the longitudinal and intrinsic transverse motion of partons inside a hadron, and the spin degrees of freedom.
In this scheme, the description of a polarised nucleon within the leading-twist approximation of perturbative QCD requires eight transverse-momentum-dependent (TMD) parton distribution functions (PDFs), which describe the distributions of longitudinal and transverse momenta of partons and their correlations with nucleon and quark spins. Of particular interest in this 3D description of the nucleon are two of these PDFs, the transversity function $\huq$ and the Sivers function $\fSiv{q}$.

For a given quark flavour $q$, $\huq$  is the analogue of the helicity distribution $\guq$ in the case of transversely polarised nucleons~\cite{Ralston:1979ys}, \textit{i.e.} the difference of the number of quarks with spin parallel to the nucleon spin and the number of quarks with spin opposite to the nucleon spin in  transversely polarised nucleons.
Together with $\guq$ and the number density $\fuq$,  $\huq$  is the only TMD PDF that survives the integration over transverse momentum.
The transversity functions of the valence $u$ and $d$ quarks allow access to the nucleon tensor charge, a fundamental property of the nucleon that can also be calculated in lattice QCD~\cite{Chen:2016utp,Bhattacharya:2016zcn}.
The transversity PDF is chiral odd and thus not directly observable in inclusive DIS.
In 1993, Collins suggested that it could be measured in SIDIS processes, where it appears coupled with another chiral-odd function~\cite{Collins:1992kk}, the so-called Collins fragmentation function $H_{1q}^{h}$.
The latter is the transverse-spin dependent fragmentation function (FF) and describes the correlation between the transverse spin of the quark $q$ and the transverse momentum of the final state hadron $h$.
This correlation leads to a left--right asymmetry in the distribution of hadrons produced in the fragmentation of transversely polarised quarks,
which in SIDIS gives rise to a transverse-spin asymmetry, the so-called Collins asymmetry $\Acoll$.
At leading order in perturbative QCD, this asymmetry
is proportional to the convolution~\cite{Kotzinian:1994dv,Mulders:1995dh,Bacchetta:2006tn} of transversity function and Collins fragmentation function over transverse momenta.
Independent essential information on $\Hhuq$ is obtained from measurements of the $e^+e^-$ annihilation process~\cite{Abe:2005zx,Seidl:2008xc,BaBar:2013jdt,BaBar:2015mcn,BESIII:2015fyw}.

\begin{figure}
\centerline{\includegraphics[width=0.7\columnwidth]{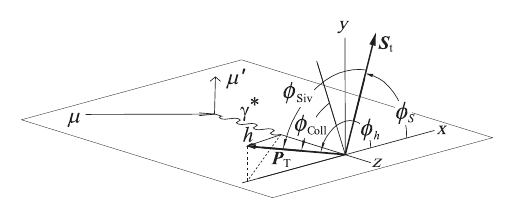}}
\caption{The SIDIS reference system defined according to
Ref.~\cite{Bacchetta:2004jz} and definition of the azimuthal angles. Here $\stv$ and $\ptv$ are the projections of the quark spin vector and of the hadron $h$ momentum vector respectively,  onto the plane transverse to the direction of the virtual photon $\gamma^*$.\label{fig:angles}}
\end{figure}

In SIDIS, the Collins effect produces a modulation
$[1 + \epsilon_\Coll \sin \phiColl]$ in the number of final state hadrons, where $\phiColl = \phi_h + \phi_S - \pi$ is the Collins angle, and $\phi_h$ and
$\phi_S$ are the azimuthal angles of the hadron and of the target nucleon spin vector, respectively, in the reference system illustrated in Fig.~\ref{fig:angles}.
The amplitude of this modulation is $\epsilon_\Coll = \Dnn f \Pt \Acoll$, where $\Dnn \simeq (1 - y)/(1 - y + y^2/2)$ is the transverse-spin transfer
coefficient from target quark to struck quark, $y$ is the fractional energy of the virtual photon, $f$ is the dilution factor of the target
material, and $\Pt$ is the transverse polarisation of the target nucleon.

The Sivers TMD PDF $\fSiv{q}$ was introduced in 1990~\cite{Sivers:1989cc} to explain the large transverse spin effects observed in $pp$ collisions.
It describes the correlation of the intrinsic transverse momentum $\kt$ of the unpolarised quarks with the transverse spin of the nucleon. The final state interactions of the struck quark with the nucleon
remnants lead to a modulation of the type $[1 + \epsilon_\Siv \sin \phiSiv]$ in distribution of the hadrons produced in SIDIS off transversely polarised nucleons.
 Here $\phiSiv = \phi_h - \phi_S$ is the Sivers angle, and the amplitude of the modulation can be written as $\epsilon_\Siv = f \Pt \Asiv$,  with $\Asiv$ being the Sivers
asymmetry~\footnote{
 Note that the  Collins and Sivers asymmetries are indicated with
 $A_{\rm{UT}}^{\sin(\phi_{\rm{h}}+\phi_{\rm{S}}-\pi)}$ and $A_{\rm{UT}}^{\sin(\phi_{\rm{h}}-\phi_{\rm{S}})}$ respectively in Ref.~\cite{COMPASS:2016led} and with $A_{\rm{UT}}^{\sin\phiColl}$ and $A_{\rm{UT}}^{\sin\phiSiv}$ in Ref.~\cite{Alexeev:2022wgr}.} that is proportional to the convolution of $\fSiv{q}$ and the spin-averaged fragmentation function $\Dhuq$.
Measuring azimuthal distributions of hadrons produced in SIDIS off a transversely polarised target allows the Collins and the Sivers effects to be disentangled~\cite{Boer:1997nt}. The Collins and Sivers asymmetries are extracted from the same data, requiring measurements with proton and deuteron
(or neutron) targets to allow for quark flavour separation.

The Collins and Sivers asymmetries were measured by the HERMES Collaboration at DESY scattering 27.6  \gevv\ electrons on transversely polarised protons~\cite{HERMES:2009lmz,Airapetian:2010ds}, by COMPASS with 160 \gevv\ muons on transversely polarised deuterons (in the years 2002--2004)~\cite{Alexakhin:2005iw,COMPASS:2006mkl,COMPASS:2008isr} and protons (in 2007 and 2010)~\cite{Alekseev:2010rw,Adolph:2012sn,COMPASS:2012dmt}, and by JLab with 5.9 \gevv\ electrons on transversely polarised $^3$He~\cite{JeffersonLabHallA:2011ayy,JeffersonLabHallA:2014yxb}.
Clear signals of non-zero asymmetries were observed on protons, while the deuteron data and the $^3$He data exhibited asymmetries compatible with zero, although with rather large statistical uncertainties.
Extractions of Sivers and transversity functions from HERMES and COMPASS data, and  $e^+e^-$ Belle data~\cite{Seidl:2008xc}, were soon carried out by several groups~\cite{Anselmino:2012rq,Anselmino:2013vqa,Martin:2014wua,Kang:2015msa,Anselmino:2015sxa,Ethier:2017zbq,Martin:2017yms}, but the largely  unbalanced statistics of the proton and deuteron data hampered an accurate extraction of the $d$-quark PDFs.
Consequently, more deuteron data were mandatory.
Eventually, in 2022 the COMPASS Collaboration took data using a transversely polarised
$^6$LiD target.

The principle of the measurements, the COMPASS apparatus and the data analysis were already described in Refs~\cite{Adolph:2012sn,COMPASS:2012dmt} and  are only briefly summarized here for completeness.
The spectrometer~\cite{Abbon:2007pq} was located in the CERN SPS North Area at the M2 beamline. In various configurations, COMPASS took data from 2002 to 2022.
In order to combine large geometrical acceptance and wide kinematic coverage, the spectrometer consists of two magnetic stages. It comprises
a variety of tracking detectors, a fast RICH, hadron and electromagnetic calorimeters, and provides muon identification via filtering through thick absorbers.
The first stage uses a 1 Tm dipole magnet with with an acceptance of about $\pm$200 mrad in both vertical and horizontal planes.
The second stage uses a 4.4 Tm spectrometer magnet, located 18 m downstream from the target, and the
acceptance is $\pm 50$ and $\pm 25$ mrad in the horizontal and vertical plane, respectively. The detector components, the front-end electronics,
the triggering system, the data acquisition and the data storage were  designed to stand the associated rate of secondaries.

In 2022, the spectrometer configuration was very similar to that used in 2007 and 2010~\cite{Abbon:2007pq}, when SIDIS off transversely polarised protons was measured.
In particular, the 180 mrad angular acceptance was significantly larger as compared to that of 70 mrad of the 2002-2004 measurements with the deuteron target.

The data were collected using a $\mu^+$  beam with a nominal momentum of 160 \gevcc, as for all COMPASS measurements to study transverse-spin effects.
The muons originated from the decay of $\pi$ and $K$ mesons produced by the 400 \gevv\ SPS proton beam on a primary beryllium target and were naturally
polarised by the weak decay mechanism. The beam polarisation was about $-80\%$ and the momentum spread was $\Delta p/p = \pm 5\%$.

The target magnet can provide both a solenoid field up to 2.5 T and a dipole field up to 0.6 T.
With such a configuration, the target polarisation can be oriented either longitudinally or transversely to the beam direction.
The target was cooled
by a $^3$He -$^4$He dilution refrigerator, reaching $\sim 50$ mK in the frozen-spin mode.
As for the previous measurement, $^6$LiD was used as a deuteron target, since its favourable dilution factor ($\sim 0.4$) and the high polarisation achievable
($\sim 0.5$) are of utmost importance.
The target consisted of three cylindrical cells with a diameter of 3 cm, mounted coaxially to the beam. The central cell was 60 cm long, and the two outer ones were 30 cm long and 5 cm apart. Neighbouring cells were polarised in opposite vertical directions, so that data for both spin directions were recorded at the same time.
The data taking was organised in periods, which were characterised by stable spectrometer performances. In the middle of each period, \textit{i.e.} after three to six days, the spin orientation in the target was reversed to further minimise systematic effects.

In this Letter we present the analysis and the results from seven out of ten periods (corresponding to about two-thirds of the total statistics collected in 2022), for which data processing and systematic studies are completed.
Candidate events are required to have reconstructed incoming and outgoing muons and reconstructed charged hadrons stemming from the muon interaction vertex.
In order to ensure the DIS regime, only events with photon virtuality $Q^2 > 1 \, \gev2$, $0.1 < y < 0.9$, and mass of the
hadronic final state system $W > 5 \, \gevc2$ are considered.
The hadrons are required to have a transverse momentum with respect to the virtual photon direction of $\pht > 0.1$  \gevcc\  and a fraction of the available energy of $z > 0.2$,
leading to a total of about $40 \times 10^6$ positive hadrons and about $32 \times 10^6$ negative hadrons.
Most of these hadrons are pions (about 70\% for positive and 75\% for negative hadrons), almost independent of the Bjorken variable $x$, of $z$ and of $\pht$~\cite{Adolph:2014zba}.

The asymmetries are measured separately for positive and negative hadrons as a function of $x$, $z$ or $\pht$.
The binning is the same as used for the deuteron results from the 2002--2004 data~\cite{COMPASS:2006mkl} and the proton results from the 2007 and 2010 data~\cite{Alekseev:2010rw,Adolph:2012sn,COMPASS:2012dmt}.
In every bin of $x$, $z$ or $\pht$, for each period the asymmetries are extracted from the number of hadrons produced in each cell for the two directions of the target polarisation.
Using an extended Unbinned Maximum Likelihood estimator~\cite{Alekseev:2010rw}, all the 8 azimuthal modulations expected in the transverse spin-dependent part of the SIDIS cross section~\cite{Bacchetta:2006tn} are fitted simultaneously.
In order to extract the Collins and Sivers asymmetries, the measured amplitudes of the modulations in $\sin \phiColl$ and $\sin \phiSiv$  are divided by
the mean values of the factors $\Dnn f \Pt$ and  $f \Pt$, respectively.
The dilution factor of the $^6$LiD material is calculated for semi-inclusive reactions as a function of $x$ and $y$~\cite{COMPASS:2005xxc}; its mean value increases with $x$
from 0.34 to 0.40, and it is $\simeq 0.38$ for all $z$ and $\pht$ bins.
The values of the deuteron target polarisation (about 0.4) are obtained from measurements, performed for each target cell and each period before and after repolarisation and taking into account the relaxation time.
The final asymmetries are obtained by averaging the results from the six periods, after verifying their statistical compatibility.

As in the previous analyses~\cite{Alekseev:2010rw, Adolph:2012sn}, detailed studies were performed in order to quantify the systematic uncertainties.
Various types of false asymmetries are measured from different combinations of the data, assuming a wrong direction of the target cell polarisation.
Differences between asymmetries extracted by splitting the data according to the reconstructed direction of the scattered muon in the spectrometer are also taken into account.
The overall systematic
point-to-point uncertainties are evaluated to be less than half of the statistical uncertainties.
Systematic scale uncertainties are  arising from the measurement of the target polarisation (3\%) and from the evaluation of the target dilution
factor (2\%).

\begin{figure*}
\centerline{\includegraphics[width=0.8\textwidth]{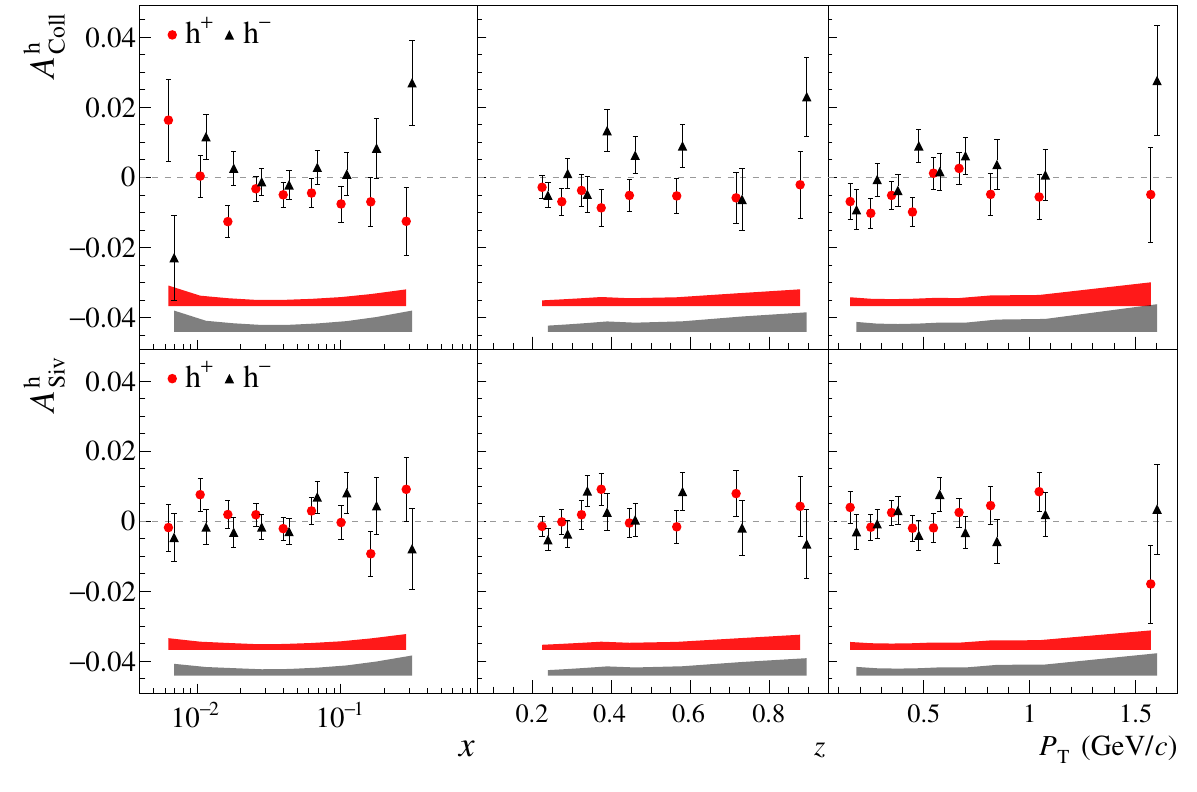}}
\caption{Results for the Collins (top) and Sivers (bottom) asymmetries for deuterons from 2022 data as a function of $x, \, z$ and $\pht$ for positive (red circles) and negative (black triangles) hadrons. The error bars are statistical only. The bands show the systematic point-to-point uncertainties.}
\label{fig:results}
\end{figure*}

Figure~\ref{fig:results} shows $\Acoll$ (top row) and $\Asiv$ (bottom row) measured as a function of $x$, $z$ and $\pht$ for positive and negative
hadrons.
The values are in good agreement with our previous measurements~\cite{Alexakhin:2005iw,COMPASS:2006mkl,COMPASS:2008isr}, with an important gain in precision: the statistical and the systematic
uncertainties are reduced by up to a factor of three.
The present measurement of $\Acoll$ indicates small negative (positive)
signals for positive (negative) hadrons at large $x$, \textit{i.e.} asymmetries with the same sign
as those measured with the proton target.
The Sivers asymmetries
are again compatible with zero, despite the much smaller
uncertainties.
All numerical values are available on HEPData~\cite{Maguire:2017ypu}.

The present results are used together with the published proton~\cite{Adolph:2012sn, COMPASS:2012dmt} and deuteron~\cite{COMPASS:2006mkl} results of COMPASS  to extract for $u$ and $d$ valence quarks the transversity distributions and the first $\ktt$-moments of the Sivers functions. The latter are defined as
$
f_{1T}^{\perp q (1)}(x, Q^2) \equiv \int \mathrm{d}^2  \kt
\, \frac{\ktt}{2 M^2} \,
\, f_{1T}^{\perp q}(x, \ktt, Q^2)$, with $M$ being the proton mass (see \textit{e.g.} Ref.~\cite{Martin:2017yms}).

The point-by-point extraction is performed by combining the proton and deuteron asymmetries in each $x$ bin, following the simple and direct procedure of Refs~\cite{Martin:2014wua, COMPASS:2018shj} and \cite{Martin:2017yms}.

\begin{figure*}%
\centerline{\includegraphics[width=0.90\textwidth]{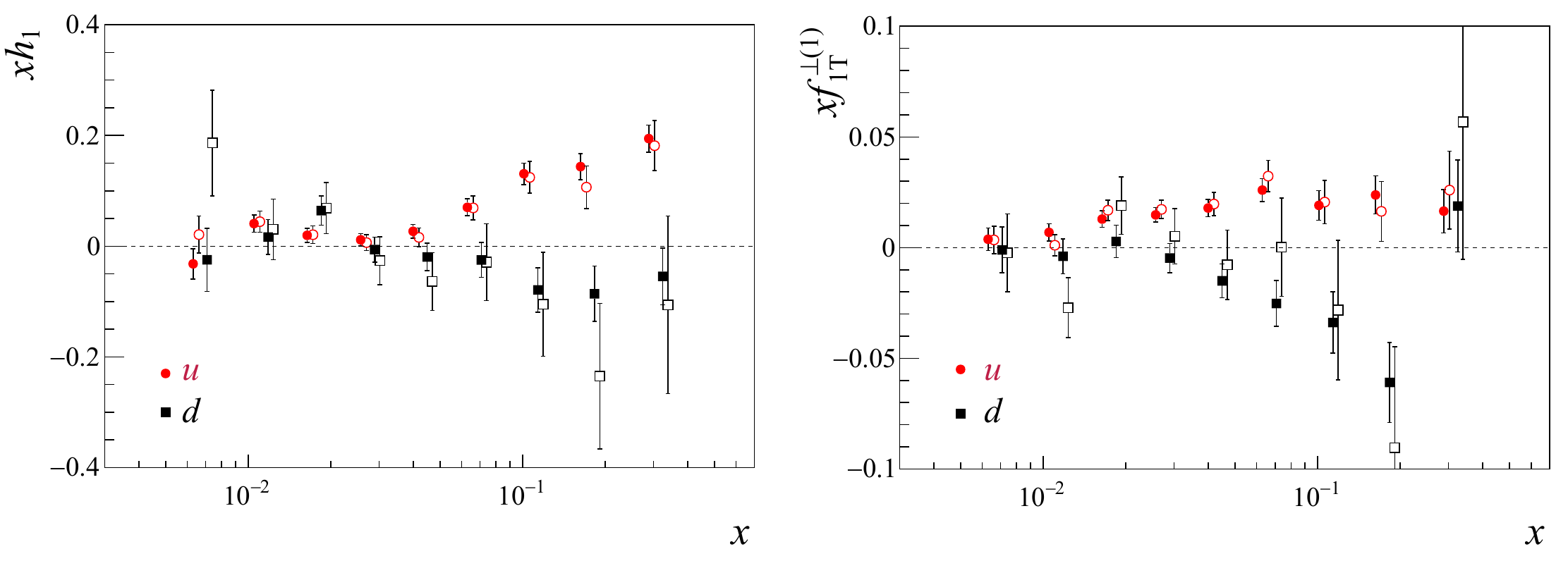} }
\caption{Left: valence transversity functions for $u$ (red circles) and $d$ (black squares) quarks. The open points show the values obtained using the previously published results for the proton and deuteron Collins asymmetries. The filled points show the values obtained including the present deuteron results. The error bars show the statistical uncertainties.
Right: the same for the first $\ktt$ moments of the Sivers functions.}
\label{fig:pdfs}
\end{figure*}

For the determination of the transversity distribution $h_1$, the
same Collins analysing power obtained from the Belle $e^+ e^- \rightarrow \rm{hadrons}$ data~\cite{Abe:2005zx,Seidl:2008xc,BaBar:2013jdt} and the same spin-averaged PDFs and FFs as in Ref.~\cite{Martin:2014wua} are used.
The results for the $u$- and $d$-valence quark are shown in the left panel of Fig.~\ref{fig:pdfs}.
The open points are the values obtained using the previously published results for $\Acoll$ and are the same as in Ref.~\cite{Martin:2014wua},
while the closed points are the values obtained using the weighted mean of the published and the present deuteron results.
A considerable reduction of the uncertainties is observed in particular for $h_1^{d_v}$, reaching almost a factor of 4 in the last  $x$ bins.
The good balance between the present proton and deuteron data strongly reduces the statistical correlation between the extracted functions $h_1^{u_v}$ and $h_1^{d_v}$ in all $x$ bins (correlation coefficient $0<\rho < 0.1$).
Table~\ref{tab:Ctensor} shows that the new COMPASS measurement clearly improves  the knowledge about the truncated tensor charges $\delta u$, $\delta d$ and $\gt$, as obtained by integrating $h_1$ over the range $0.008<x<0.210$.
For the $u$ quark, new and old values are consistent, with a reduction of the statistical uncertainty by about 30\% when adding the present results. For the $d$ quark, the new values are a factor of about 2.5 smaller, with a reduction of the statistical uncertainty by a factor of two. The isovector nucleon tensor charge $\gt$ is now about 20\% smaller, with the uncertainty reduced by almost a factor of two.
\begin{table}
\begin{center}
\begin{tabular}{|l|c|c|c|c|c|c|}
\hline
data & $\delta u =\int_{0.008}^{0.210} \mathrm{d}x\, h_1^{u_v}(x)$ & $\delta d = \int_{0.008}^{0.210} \mathrm{d}x\, h_1^{d_v}(x)$ & $\gt = \delta u - \delta d$ \bigstrut\\ \hline
previous~\cite{COMPASS:2006mkl,Adolph:2012sn,COMPASS:2012dmt}       &    0.187 $\pm$ 0.030 & -0.178 $\pm$ 0.097 &  0.365 $\pm$ 0.078 \bigstrut\\ \hline
previous~~\cite{COMPASS:2006mkl,Adolph:2012sn,COMPASS:2012dmt} and present   &    0.214 $\pm$ 0.020 & -0.070 $\pm$ 0.043 &  0.284 $\pm$ 0.045 \bigstrut\\ \hline
\end{tabular}
\end{center}
\caption{Truncated tensor charges $\delta u$, $\delta d$ and $\gt$ in the range $0.008<x<0.210$ from
numerical integration of the $h_1$ values, obtained using for $\Acoll$ the previously published results only (first line) and adding the present results (second line).}
\label{tab:Ctensor}
\end{table}

For the Sivers function, we follow the procedure of Ref.~\cite{Martin:2017yms} using the same spin-averaged PDFs and FFs and assuming that all charged hadrons are pions.
The right panel of Fig.~\ref{fig:pdfs} shows the results for $\mfSiv{\, u_v}$ and $\mfSiv{\, d_v}$.
The open points are the values obtained from the previously published $\Asiv$ results only, while the filled ones are obtained adding the present deuteron results.
For the $d$ quark, the statistical uncertainties are reduced by about a factor of two, and the different $x$ dependence for $u$ and $d$ quarks is now quite clear.

In summary, this Letter presents the new high-statistics COMPASS results for the Collins and the Sivers asymmetries of charged hadrons measured
from part of the SIDIS data collected in 2022 with a deuteron target. The asymmetries are given as a function of $x$, $z$ or $\pht$.
They are consistent with the previously published deuteron results, and the overall precision is improved by up to a factor of three. Combining the new deuteron data with all proton and deuteron data from COMPASS, both the transversity distribution and the first moments of the Sivers function are extracted. By numerically integrating the transversity distributions of the $u$ and $d$ valence quarks the tensor charge is evaluated in the range  $0.008 < x < 0.21$.
Altogether, the new COMPASS deuteron results are expected to have strong impact on the knowledge of the transverse-spin structure of the nucleon.

{\bf Acknowledgements}\\
We acknowledge the support of the CERN management and staff, as well as the skills and efforts of the technicians of the collaborating institutes. In particular, we are grateful to the CERN Management and SPSC  for supporting the completion of COMPASS data taking with a full run in 2022.
We thank Vincenzo Barone and Gunar Schnell for useful discussions.


\begin{thebibliography}{43}%
\makeatletter
\providecommand \@ifxundefined [1]{%
 \@ifx{#1\undefined}
}%
\providecommand \@ifnum [1]{%
 \ifnum #1\expandafter \@firstoftwo
 \else \expandafter \@secondoftwo
 \fi
}%
\providecommand \@ifx [1]{%
 \ifx #1\expandafter \@firstoftwo
 \else \expandafter \@secondoftwo
 \fi
}%
\providecommand \natexlab [1]{#1}%
\providecommand \enquote  [1]{``#1''}%
\providecommand \bibnamefont  [1]{#1}%
\providecommand \bibfnamefont [1]{#1}%
\providecommand \citenamefont [1]{#1}%
\providecommand \href@noop [0]{\@secondoftwo}%
\providecommand \href [0]{\begingroup \@sanitize@url \@href}%
\providecommand \@href[1]{\@@startlink{#1}\@@href}%
\providecommand \@@href[1]{\endgroup#1\@@endlink}%
\providecommand \@sanitize@url [0]{\catcode `\\12\catcode `\$12\catcode
  `\&12\catcode `\#12\catcode `\^12\catcode `\_12\catcode `\%12\relax}%
\providecommand \@@startlink[1]{}%
\providecommand \@@endlink[0]{}%
\providecommand \url  [0]{\begingroup\@sanitize@url \@url }%
\providecommand \@url [1]{\endgroup\@href {#1}{\urlprefix }}%
\providecommand \urlprefix  [0]{URL }%
\providecommand \Eprint [0]{\href }%
\providecommand \doibase [0]{https://doi.org/}%
\providecommand \selectlanguage [0]{\@gobble}%
\providecommand \bibinfo  [0]{\@secondoftwo}%
\providecommand \bibfield  [0]{\@secondoftwo}%
\providecommand \translation [1]{[#1]}%
\providecommand \BibitemOpen [0]{}%
\providecommand \bibitemStop [0]{}%
\providecommand \bibitemNoStop [0]{.\EOS\space}%
\providecommand \EOS [0]{\spacefactor3000\relax}%
\providecommand \BibitemShut  [1]{\csname bibitem#1\endcsname}%
\let\auto@bib@innerbib\@empty
\bibitem [{\citenamefont {Bunce}\ \emph {et~al.}(1976)\citenamefont {Bunce}
  \emph {et~al.}}]{Bunce:1976yb}%
  \BibitemOpen
  \bibfield  {author} {\bibinfo {author} {\bibfnamefont {G.}~\bibnamefont
  {Bunce}} \emph {et~al.},\ }\href
  {https://doi.org/10.1103/PhysRevLett.36.1113} {\bibfield  {journal} {\bibinfo
   {journal} {Phys. Rev. Lett.}\ }\textbf {\bibinfo {volume} {36}},\ \bibinfo
  {pages} {1113} (\bibinfo {year} {1976})}\BibitemShut {NoStop}%
\bibitem [{\citenamefont {Kane}\ \emph {et~al.}(1978)\citenamefont {Kane},
  \citenamefont {Pumplin},\ and\ \citenamefont {Repko}}]{Kane:1978nd}%
  \BibitemOpen
  \bibfield  {author} {\bibinfo {author} {\bibfnamefont {G.~L.}\ \bibnamefont
  {Kane}}, \bibinfo {author} {\bibfnamefont {J.}~\bibnamefont {Pumplin}},\ and\
  \bibinfo {author} {\bibfnamefont {W.}~\bibnamefont {Repko}},\ }\href
  {https://doi.org/10.1103/PhysRevLett.41.1689} {\bibfield  {journal} {\bibinfo
   {journal} {Phys. Rev. Lett.}\ }\textbf {\bibinfo {volume} {41}},\ \bibinfo
  {pages} {1689} (\bibinfo {year} {1978})}\BibitemShut {NoStop}%
\bibitem [{\citenamefont {Ralston}\ and\ \citenamefont
  {Soper}(1979)}]{Ralston:1979ys}%
  \BibitemOpen
  \bibfield  {author} {\bibinfo {author} {\bibfnamefont {J.~P.}\ \bibnamefont
  {Ralston}}\ and\ \bibinfo {author} {\bibfnamefont {D.~E.}\ \bibnamefont
  {Soper}},\ }\href {https://doi.org/10.1016/0550-3213(79)90082-8} {\bibfield
  {journal} {\bibinfo  {journal} {Nucl. Phys. B}\ }\textbf {\bibinfo {volume}
  {152}},\ \bibinfo {pages} {109} (\bibinfo {year} {1979})}\BibitemShut
  {NoStop}%
\bibitem [{\citenamefont {Artru}\ and\ \citenamefont
  {Mekhfi}(1990)}]{Artru:1989zv}%
  \BibitemOpen
  \bibfield  {author} {\bibinfo {author} {\bibfnamefont {X.}~\bibnamefont
  {Artru}}\ and\ \bibinfo {author} {\bibfnamefont {M.}~\bibnamefont {Mekhfi}},\
  }\href {https://doi.org/10.1007/BF01556280} {\bibfield  {journal} {\bibinfo
  {journal} {Z. Phys. C}\ }\textbf {\bibinfo {volume} {45}},\ \bibinfo {pages}
  {669} (\bibinfo {year} {1990})}\BibitemShut {NoStop}%
\bibitem [{\citenamefont {Jaffe}\ and\ \citenamefont
  {Ji}(1991)}]{Jaffe:1991kp}%
  \BibitemOpen
  \bibfield  {author} {\bibinfo {author} {\bibfnamefont {R.~L.}\ \bibnamefont
  {Jaffe}}\ and\ \bibinfo {author} {\bibfnamefont {X.}~\bibnamefont {Ji}},\
  }\href {https://doi.org/10.1103/PhysRevLett.67.552} {\bibfield  {journal}
  {\bibinfo  {journal} {Phys. Rev. Lett.}\ }\textbf {\bibinfo {volume} {67}},\
  \bibinfo {pages} {552} (\bibinfo {year} {1991})}\BibitemShut {NoStop}%
\bibitem [{\citenamefont {Bacchetta}\ \emph {et~al.}(2007)\citenamefont
  {Bacchetta} \emph {et~al.}}]{Bacchetta:2006tn}%
  \BibitemOpen
  \bibfield  {author} {\bibinfo {author} {\bibfnamefont {A.}~\bibnamefont
  {Bacchetta}} \emph {et~al.},\ }\href
  {https://doi.org/10.1088/1126-6708/2007/02/093} {\bibfield  {journal}
  {\bibinfo  {journal} {JHEP}\ }\textbf {\bibinfo {volume} {0702}},\ \bibinfo
  {pages} {093}},\ \Eprint {https://arxiv.org/abs/hep-ph/0611265}
  {arXiv:hep-ph/0611265 [hep-ph]} \BibitemShut {NoStop}%
\bibitem [{\citenamefont {Chen}\ \emph {et~al.}(2016)\citenamefont {Chen},
  \citenamefont {Cohen}, \citenamefont {Ji}, \citenamefont {Lin},\ and\
  \citenamefont {Zhang}}]{Chen:2016utp}%
  \BibitemOpen
  \bibfield  {author} {\bibinfo {author} {\bibfnamefont {J.-W.}\ \bibnamefont
  {Chen}}, \bibinfo {author} {\bibfnamefont {S.~D.}\ \bibnamefont {Cohen}},
  \bibinfo {author} {\bibfnamefont {X.}~\bibnamefont {Ji}}, \bibinfo {author}
  {\bibfnamefont {H.-W.}\ \bibnamefont {Lin}},\ and\ \bibinfo {author}
  {\bibfnamefont {J.-H.}\ \bibnamefont {Zhang}},\ }\href
  {https://doi.org/10.1016/j.nuclphysb.2016.07.033} {\bibfield  {journal}
  {\bibinfo  {journal} {Nucl. Phys. B}\ }\textbf {\bibinfo {volume} {911}},\
  \bibinfo {pages} {246} (\bibinfo {year} {2016})},\ \Eprint
  {https://arxiv.org/abs/1603.06664} {arXiv:1603.06664 [hep-ph]} \BibitemShut
  {NoStop}%
\bibitem [{\citenamefont {Bhattacharya}\ \emph {et~al.}(2016)\citenamefont
  {Bhattacharya} \emph {et~al.}}]{Bhattacharya:2016zcn}%
  \BibitemOpen
  \bibfield  {author} {\bibinfo {author} {\bibfnamefont {T.}~\bibnamefont
  {Bhattacharya}} \emph {et~al.},\ }\href
  {https://doi.org/10.1103/PhysRevD.94.054508} {\bibfield  {journal} {\bibinfo
  {journal} {Phys. Rev. D}\ }\textbf {\bibinfo {volume} {94}},\ \bibinfo
  {pages} {054508} (\bibinfo {year} {2016})},\ \Eprint
  {https://arxiv.org/abs/1606.07049} {arXiv:1606.07049 [hep-lat]} \BibitemShut
  {NoStop}%
\bibitem [{\citenamefont {Collins}(1993)}]{Collins:1992kk}%
  \BibitemOpen
  \bibfield  {author} {\bibinfo {author} {\bibfnamefont {J.~C.}\ \bibnamefont
  {Collins}},\ }\href {https://doi.org/10.1016/0550-3213(93)90262-N} {\bibfield
   {journal} {\bibinfo  {journal} {Nucl. Phys. B}\ }\textbf {\bibinfo {volume}
  {396}},\ \bibinfo {pages} {161} (\bibinfo {year} {1993})},\ \Eprint
  {https://arxiv.org/abs/hep-ph/9208213} {arXiv:hep-ph/9208213} \BibitemShut
  {NoStop}%
\bibitem [{\citenamefont {{A. Kotzinian,}}(1995)}]{Kotzinian:1994dv}%
  \BibitemOpen
  \bibfield  {author} {\bibinfo {author} {\bibnamefont {{A. Kotzinian,}}},\
  }\href {https://doi.org/10.1016/0550-3213(95)00098-D} {\bibfield  {journal}
  {\bibinfo  {journal} {Nucl. Phys. B}\ }\textbf {\bibinfo {volume} {441}},\
  \bibinfo {pages} {234} (\bibinfo {year} {1995})},\ \Eprint
  {https://arxiv.org/abs/hep-ph/9412283} {arXiv:hep-ph/9412283} \BibitemShut
  {NoStop}%
\bibitem [{\citenamefont {Mulders}\ and\ \citenamefont
  {Tangerman}(1996)}]{Mulders:1995dh}%
  \BibitemOpen
  \bibfield  {author} {\bibinfo {author} {\bibfnamefont {P.~J.}\ \bibnamefont
  {Mulders}}\ and\ \bibinfo {author} {\bibfnamefont {R.~D.}\ \bibnamefont
  {Tangerman}},\ }\href {https://doi.org/10.1016/0550-3213(95)00632-X}
  {\bibfield  {journal} {\bibinfo  {journal} {Nucl. Phys. B}\ }\textbf
  {\bibinfo {volume} {461}},\ \bibinfo {pages} {197} (\bibinfo {year}
  {1996})},\ \bibinfo {note} {[Erratum-ibid.\ {\it B} {\bf 484} (1997) 538 ]},\
  \Eprint {https://arxiv.org/abs/hep-ph/9510301} {arXiv:hep-ph/9510301}
  \BibitemShut {NoStop}%
\bibitem [{\citenamefont {Abe}\ \emph {et~al.}(2006)\citenamefont {Abe} \emph
  {et~al.}}]{Abe:2005zx}%
  \BibitemOpen
  \bibfield  {author} {\bibinfo {author} {\bibfnamefont {K.}~\bibnamefont
  {Abe}} \emph {et~al.} (\bibinfo {collaboration} {Belle collaboration}),\
  }\href {https://doi.org/10.1103/PhysRevLett.96.232002} {\bibfield  {journal}
  {\bibinfo  {journal} {Phys. Rev. Lett.}\ }\textbf {\bibinfo {volume} {96}},\
  \bibinfo {pages} {232002} (\bibinfo {year} {2006})},\ \Eprint
  {https://arxiv.org/abs/hep-ex/0507063} {arXiv:hep-ex/0507063} \BibitemShut
  {NoStop}%
\bibitem [{\citenamefont {Seidl}\ \emph {et~al.}(2008)\citenamefont {Seidl}
  \emph {et~al.}}]{Seidl:2008xc}%
  \BibitemOpen
  \bibfield  {author} {\bibinfo {author} {\bibfnamefont {R.}~\bibnamefont
  {Seidl}} \emph {et~al.} (\bibinfo {collaboration} {Belle collaboration}),\
  }\href {https://doi.org/10.1103/PhysRevD.78.032011} {\bibfield  {journal}
  {\bibinfo  {journal} {Phys. Rev. D}\ }\textbf {\bibinfo {volume} {78}},\
  \bibinfo {pages} {032011} (\bibinfo {year} {2008})},\ \Eprint
  {https://arxiv.org/abs/0805.2975} {arXiv:0805.2975 [hep-ex]} \BibitemShut
  {NoStop}%
\bibitem [{\citenamefont {Lees}\ \emph {et~al.}(2014)\citenamefont {Lees} \emph
  {et~al.}}]{BaBar:2013jdt}%
  \BibitemOpen
  \bibfield  {author} {\bibinfo {author} {\bibfnamefont {J.~P.}\ \bibnamefont
  {Lees}} \emph {et~al.} (\bibinfo {collaboration} {BaBar collaboration}),\
  }\href {https://doi.org/10.1103/PhysRevD.90.052003} {\bibfield  {journal}
  {\bibinfo  {journal} {Phys. Rev. D}\ }\textbf {\bibinfo {volume} {90}},\
  \bibinfo {pages} {052003} (\bibinfo {year} {2014})},\ \Eprint
  {https://arxiv.org/abs/1309.5278} {arXiv:1309.5278 [hep-ex]} \BibitemShut
  {NoStop}%
\bibitem [{\citenamefont {Lees}\ \emph {et~al.}(2015)\citenamefont {Lees} \emph
  {et~al.}}]{BaBar:2015mcn}%
  \BibitemOpen
  \bibfield  {author} {\bibinfo {author} {\bibfnamefont {J.~P.}\ \bibnamefont
  {Lees}} \emph {et~al.} (\bibinfo {collaboration} {BaBar collaboration}),\
  }\href {https://doi.org/10.1103/PhysRevD.92.111101} {\bibfield  {journal}
  {\bibinfo  {journal} {Phys. Rev. D}\ }\textbf {\bibinfo {volume} {92}},\
  \bibinfo {pages} {111101} (\bibinfo {year} {2015})},\ \Eprint
  {https://arxiv.org/abs/1506.05864} {arXiv:1506.05864 [hep-ex]} \BibitemShut
  {NoStop}%
\bibitem [{\citenamefont {Ablikim}\ \emph {et~al.}(2016)\citenamefont {Ablikim}
  \emph {et~al.}}]{BESIII:2015fyw}%
  \BibitemOpen
  \bibfield  {author} {\bibinfo {author} {\bibfnamefont {M.}~\bibnamefont
  {Ablikim}} \emph {et~al.} (\bibinfo {collaboration} {BESIII collaboration}),\
  }\href {https://doi.org/10.1103/PhysRevLett.116.042001} {\bibfield  {journal}
  {\bibinfo  {journal} {Phys. Rev. Lett.}\ }\textbf {\bibinfo {volume} {116}},\
  \bibinfo {pages} {042001} (\bibinfo {year} {2016})},\ \Eprint
  {https://arxiv.org/abs/1507.06824} {arXiv:1507.06824 [hep-ex]} \BibitemShut
  {NoStop}%
\bibitem [{\citenamefont {Bacchetta}\ \emph {et~al.}(2004)\citenamefont
  {Bacchetta}, \citenamefont {D'Alesio}, \citenamefont {Diehl},\ and\
  \citenamefont {Miller}}]{Bacchetta:2004jz}%
  \BibitemOpen
  \bibfield  {author} {\bibinfo {author} {\bibfnamefont {A.}~\bibnamefont
  {Bacchetta}}, \bibinfo {author} {\bibfnamefont {U.}~\bibnamefont {D'Alesio}},
  \bibinfo {author} {\bibfnamefont {M.}~\bibnamefont {Diehl}},\ and\ \bibinfo
  {author} {\bibfnamefont {C.~A.}\ \bibnamefont {Miller}},\ }\href
  {https://doi.org/10.1103/PhysRevD.70.117504} {\bibfield  {journal} {\bibinfo
  {journal} {Phys. Rev. D}\ }\textbf {\bibinfo {volume} {70}},\ \bibinfo
  {pages} {117504} (\bibinfo {year} {2004})},\ \Eprint
  {https://arxiv.org/abs/hep-ph/0410050} {arXiv:hep-ph/0410050} \BibitemShut
  {NoStop}%
\bibitem [{\citenamefont {Sivers}(1990)}]{Sivers:1989cc}%
  \BibitemOpen
  \bibfield  {author} {\bibinfo {author} {\bibfnamefont {D.~W.}\ \bibnamefont
  {Sivers}},\ }\href {https://doi.org/10.1103/PhysRevD.41.83} {\bibfield
  {journal} {\bibinfo  {journal} {Phys. Rev. D}\ }\textbf {\bibinfo {volume}
  {41}},\ \bibinfo {pages} {83} (\bibinfo {year} {1990})}\BibitemShut {NoStop}%
\bibitem [{\citenamefont {Adolph}\ \emph {et~al.}(2017)\citenamefont {Adolph}
  \emph {et~al.}}]{COMPASS:2016led}%
  \BibitemOpen
  \bibfield  {author} {\bibinfo {author} {\bibfnamefont {C.}~\bibnamefont
  {Adolph}} \emph {et~al.} (\bibinfo {collaboration} {COMPASS collaboration}),\
  }\href {https://doi.org/10.1016/j.physletb.2017.04.042} {\bibfield  {journal}
  {\bibinfo  {journal} {Phys. Lett. B}\ }\textbf {\bibinfo {volume} {770}},\
  \bibinfo {pages} {138} (\bibinfo {year} {2017})},\ \Eprint
  {https://arxiv.org/abs/1609.07374} {arXiv:1609.07374 [hep-ex]} \BibitemShut
  {NoStop}%
\bibitem [{\citenamefont {Alexeev}\ \emph {et~al.}(2023)\citenamefont {Alexeev}
  \emph {et~al.}}]{Alexeev:2022wgr}%
  \BibitemOpen
  \bibfield  {author} {\bibinfo {author} {\bibfnamefont {G.~D.}\ \bibnamefont
  {Alexeev}} \emph {et~al.} (\bibinfo {collaboration} {COMPASS
  collaboration}),\ }\href {https://doi.org/10.1016/j.physletb.2023.137950}
  {\bibfield  {journal} {\bibinfo  {journal} {Phys. Lett. B}\ }\textbf
  {\bibinfo {volume} {843}},\ \bibinfo {pages} {137950} (\bibinfo {year}
  {2023})},\ \Eprint {https://arxiv.org/abs/2211.00093} {arXiv:2211.00093
  [hep-ex]} \BibitemShut {NoStop}%
\bibitem [{\citenamefont {Boer}\ and\ \citenamefont
  {Mulders}(1998)}]{Boer:1997nt}%
  \BibitemOpen
  \bibfield  {author} {\bibinfo {author} {\bibfnamefont {D.}~\bibnamefont
  {Boer}}\ and\ \bibinfo {author} {\bibfnamefont {P.~J.}\ \bibnamefont
  {Mulders}},\ }\href {https://doi.org/10.1103/PhysRevD.57.5780} {\bibfield
  {journal} {\bibinfo  {journal} {Phys. Rev. D}\ }\textbf {\bibinfo {volume}
  {57}},\ \bibinfo {pages} {5780} (\bibinfo {year} {1998})},\ \Eprint
  {https://arxiv.org/abs/hep-ph/9711485} {arXiv:hep-ph/9711485} \BibitemShut
  {NoStop}%
\bibitem [{\citenamefont {Airapetian}\ \emph {et~al.}(2009)\citenamefont
  {Airapetian} \emph {et~al.}}]{HERMES:2009lmz}%
  \BibitemOpen
  \bibfield  {author} {\bibinfo {author} {\bibfnamefont {A.}~\bibnamefont
  {Airapetian}} \emph {et~al.} (\bibinfo {collaboration} {HERMES
  collaboration}),\ }\href {https://doi.org/10.1103/PhysRevLett.103.152002}
  {\bibfield  {journal} {\bibinfo  {journal} {Phys. Rev. Lett.}\ }\textbf
  {\bibinfo {volume} {103}},\ \bibinfo {pages} {152002} (\bibinfo {year}
  {2009})},\ \Eprint {https://arxiv.org/abs/0906.3918} {arXiv:0906.3918
  [hep-ex]} \BibitemShut {NoStop}%
\bibitem [{\citenamefont {Airapetian}\ \emph {et~al.}(2010)\citenamefont
  {Airapetian} \emph {et~al.}}]{Airapetian:2010ds}%
  \BibitemOpen
  \bibfield  {author} {\bibinfo {author} {\bibfnamefont {A.}~\bibnamefont
  {Airapetian}} \emph {et~al.} (\bibinfo {collaboration} {HERMES
  collaboration}),\ }\href {https://doi.org/10.1016/j.physletb.2010.08.012}
  {\bibfield  {journal} {\bibinfo  {journal} {Phys. Lett. B}\ }\textbf
  {\bibinfo {volume} {693}},\ \bibinfo {pages} {11} (\bibinfo {year} {2010})},\
  \Eprint {https://arxiv.org/abs/1006.4221} {arXiv:1006.4221 [hep-ex]}
  \BibitemShut {NoStop}%
\bibitem [{\citenamefont {Alexakhin}\ \emph {et~al.}(2005)\citenamefont
  {Alexakhin} \emph {et~al.}}]{Alexakhin:2005iw}%
  \BibitemOpen
  \bibfield  {author} {\bibinfo {author} {\bibfnamefont {V.}~\bibnamefont
  {Alexakhin}} \emph {et~al.} (\bibinfo {collaboration} {COMPASS
  collaboration}),\ }\href {https://doi.org/10.1103/PhysRevLett.94.202002}
  {\bibfield  {journal} {\bibinfo  {journal} {Phys. Rev. Lett.}\ }\textbf
  {\bibinfo {volume} {94}},\ \bibinfo {pages} {202002} (\bibinfo {year}
  {2005})},\ \Eprint {https://arxiv.org/abs/hep-ex/0503002}
  {arXiv:hep-ex/0503002} \BibitemShut {NoStop}%
\bibitem [{\citenamefont {Ageev}\ \emph {et~al.}(2007)\citenamefont {Ageev}
  \emph {et~al.}}]{COMPASS:2006mkl}%
  \BibitemOpen
  \bibfield  {author} {\bibinfo {author} {\bibfnamefont {E.~S.}\ \bibnamefont
  {Ageev}} \emph {et~al.} (\bibinfo {collaboration} {COMPASS collaboration}),\
  }\href {https://doi.org/10.1016/j.nuclphysb.2006.10.027} {\bibfield
  {journal} {\bibinfo  {journal} {Nucl. Phys. B}\ }\textbf {\bibinfo {volume}
  {765}},\ \bibinfo {pages} {31} (\bibinfo {year} {2007})},\ \Eprint
  {https://arxiv.org/abs/hep-ex/0610068} {arXiv:hep-ex/0610068} \BibitemShut
  {NoStop}%
\bibitem [{\citenamefont {Alekseev}\ \emph {et~al.}(2009)\citenamefont
  {Alekseev} \emph {et~al.}}]{COMPASS:2008isr}%
  \BibitemOpen
  \bibfield  {author} {\bibinfo {author} {\bibfnamefont {M.}~\bibnamefont
  {Alekseev}} \emph {et~al.} (\bibinfo {collaboration} {COMPASS
  collaboration}),\ }\href {https://doi.org/10.1016/j.physletb.2009.01.060}
  {\bibfield  {journal} {\bibinfo  {journal} {Phys. Lett. B}\ }\textbf
  {\bibinfo {volume} {673}},\ \bibinfo {pages} {127} (\bibinfo {year}
  {2009})},\ \Eprint {https://arxiv.org/abs/0802.2160} {arXiv:0802.2160
  [hep-ex]} \BibitemShut {NoStop}%
\bibitem [{\citenamefont {Alekseev}\ \emph {et~al.}(2010)\citenamefont
  {Alekseev} \emph {et~al.}}]{Alekseev:2010rw}%
  \BibitemOpen
  \bibfield  {author} {\bibinfo {author} {\bibfnamefont {M.}~\bibnamefont
  {Alekseev}} \emph {et~al.} (\bibinfo {collaboration} {COMPASS
  collaboration}),\ }\href {https://doi.org/10.1016/j.physletb.2010.08.001}
  {\bibfield  {journal} {\bibinfo  {journal} {Phys. Lett. B}\ }\textbf
  {\bibinfo {volume} {692}},\ \bibinfo {pages} {240} (\bibinfo {year}
  {2010})},\ \Eprint {https://arxiv.org/abs/1005.5609} {arXiv:1005.5609
  [hep-ex]} \BibitemShut {NoStop}%
\bibitem [{\citenamefont {Adolph}\ \emph
  {et~al.}(2012{\natexlab{a}})\citenamefont {Adolph} \emph
  {et~al.}}]{Adolph:2012sn}%
  \BibitemOpen
  \bibfield  {author} {\bibinfo {author} {\bibfnamefont {C.}~\bibnamefont
  {Adolph}} \emph {et~al.} (\bibinfo {collaboration} {COMPASS collaboration}),\
  }\href {https://doi.org/10.1016/j.physletb.2012.09.055} {\bibfield  {journal}
  {\bibinfo  {journal} {Phys. Lett. B}\ }\textbf {\bibinfo {volume} {717}},\
  \bibinfo {pages} {376} (\bibinfo {year} {2012}{\natexlab{a}})},\ \Eprint
  {https://arxiv.org/abs/1205.5121} {arXiv:1205.5121 [hep-ex]} \BibitemShut
  {NoStop}%
\bibitem [{\citenamefont {Adolph}\ \emph
  {et~al.}(2012{\natexlab{b}})\citenamefont {Adolph} \emph
  {et~al.}}]{COMPASS:2012dmt}%
  \BibitemOpen
  \bibfield  {author} {\bibinfo {author} {\bibfnamefont {C.}~\bibnamefont
  {Adolph}} \emph {et~al.} (\bibinfo {collaboration} {COMPASS collaboration}),\
  }\href {https://doi.org/10.1016/j.physletb.2012.09.056} {\bibfield  {journal}
  {\bibinfo  {journal} {Phys. Lett. B}\ }\textbf {\bibinfo {volume} {717}},\
  \bibinfo {pages} {383} (\bibinfo {year} {2012}{\natexlab{b}})},\ \Eprint
  {https://arxiv.org/abs/1205.5122} {arXiv:1205.5122 [hep-ex]} \BibitemShut
  {NoStop}%
\bibitem [{\citenamefont {Qian}\ \emph {et~al.}(2011)\citenamefont {Qian} \emph
  {et~al.}}]{JeffersonLabHallA:2011ayy}%
  \BibitemOpen
  \bibfield  {author} {\bibinfo {author} {\bibfnamefont {X.}~\bibnamefont
  {Qian}} \emph {et~al.} (\bibinfo {collaboration} {Jefferson Lab Hall A
  collaboration}),\ }\href {https://doi.org/10.1103/PhysRevLett.107.072003}
  {\bibfield  {journal} {\bibinfo  {journal} {Phys. Rev. Lett.}\ }\textbf
  {\bibinfo {volume} {107}},\ \bibinfo {pages} {072003} (\bibinfo {year}
  {2011})},\ \Eprint {https://arxiv.org/abs/1106.0363} {arXiv:1106.0363
  [nucl-ex]} \BibitemShut {NoStop}%
\bibitem [{\citenamefont {Zhao}\ \emph {et~al.}(2014)\citenamefont {Zhao} \emph
  {et~al.}}]{JeffersonLabHallA:2014yxb}%
  \BibitemOpen
  \bibfield  {author} {\bibinfo {author} {\bibfnamefont {Y.~X.}\ \bibnamefont
  {Zhao}} \emph {et~al.} (\bibinfo {collaboration} {Jefferson Lab Hall A
  collaboration}),\ }\href {https://doi.org/10.1103/PhysRevC.90.055201}
  {\bibfield  {journal} {\bibinfo  {journal} {Phys. Rev. C}\ }\textbf {\bibinfo
  {volume} {90}},\ \bibinfo {pages} {055201} (\bibinfo {year} {2014})},\
  \Eprint {https://arxiv.org/abs/1404.7204} {arXiv:1404.7204 [nucl-ex]}
  \BibitemShut {NoStop}%
\bibitem [{\citenamefont {Anselmino}\ \emph {et~al.}(2012)\citenamefont
  {Anselmino} \emph {et~al.}}]{Anselmino:2012rq}%
  \BibitemOpen
  \bibfield  {author} {\bibinfo {author} {\bibfnamefont {M.}~\bibnamefont
  {Anselmino}} \emph {et~al.},\ }\href
  {https://doi.org/10.1103/PhysRevD.86.074032} {\bibfield  {journal} {\bibinfo
  {journal} {Phys. Rev. D}\ }\textbf {\bibinfo {volume} {86}},\ \bibinfo
  {pages} {074032} (\bibinfo {year} {2012})},\ \Eprint
  {https://arxiv.org/abs/1207.6529} {arXiv:1207.6529 [hep-ph]} \BibitemShut
  {NoStop}%
\bibitem [{\citenamefont {Anselmino}\ \emph {et~al.}(2013)\citenamefont
  {Anselmino} \emph {et~al.}}]{Anselmino:2013vqa}%
  \BibitemOpen
  \bibfield  {author} {\bibinfo {author} {\bibfnamefont {M.}~\bibnamefont
  {Anselmino}} \emph {et~al.},\ }\href
  {https://doi.org/10.1103/PhysRevD.87.094019} {\bibfield  {journal} {\bibinfo
  {journal} {Phys.Rev. D}\ }\textbf {\bibinfo {volume} {87}},\ \bibinfo {pages}
  {094019} (\bibinfo {year} {2013})},\ \Eprint
  {https://arxiv.org/abs/arXiv:hep-ph/1303.3822} {arXiv:arXiv:hep-ph/1303.3822
  [hep-ph]} \BibitemShut {NoStop}%
\bibitem [{\citenamefont {Martin}\ \emph {et~al.}(2015)\citenamefont {Martin},
  \citenamefont {Bradamante},\ and\ \citenamefont {Barone}}]{Martin:2014wua}%
  \BibitemOpen
  \bibfield  {author} {\bibinfo {author} {\bibfnamefont {A.}~\bibnamefont
  {Martin}}, \bibinfo {author} {\bibfnamefont {F.}~\bibnamefont {Bradamante}},\
  and\ \bibinfo {author} {\bibfnamefont {V.}~\bibnamefont {Barone}},\ }\href
  {https://doi.org/10.1103/PhysRevD.91.014034} {\bibfield  {journal} {\bibinfo
  {journal} {Phys. Rev. D}\ }\textbf {\bibinfo {volume} {91}},\ \bibinfo
  {pages} {014034} (\bibinfo {year} {2015})},\ \Eprint
  {https://arxiv.org/abs/1412.5946} {arXiv:1412.5946 [hep-ph]} \BibitemShut
  {NoStop}%
\bibitem [{\citenamefont {Kang}\ \emph {et~al.}(2016)\citenamefont {Kang},
  \citenamefont {Prokudin}, \citenamefont {Sun},\ and\ \citenamefont
  {Yuan}}]{Kang:2015msa}%
  \BibitemOpen
  \bibfield  {author} {\bibinfo {author} {\bibfnamefont {Z.-B.}\ \bibnamefont
  {Kang}}, \bibinfo {author} {\bibfnamefont {A.}~\bibnamefont {Prokudin}},
  \bibinfo {author} {\bibfnamefont {P.}~\bibnamefont {Sun}},\ and\ \bibinfo
  {author} {\bibfnamefont {F.}~\bibnamefont {Yuan}},\ }\href
  {https://doi.org/10.1103/PhysRevD.93.014009} {\bibfield  {journal} {\bibinfo
  {journal} {Phys. Rev. D}\ }\textbf {\bibinfo {volume} {93}},\ \bibinfo
  {pages} {014009} (\bibinfo {year} {2016})},\ \Eprint
  {https://arxiv.org/abs/1505.05589} {arXiv:1505.05589 [hep-ph]} \BibitemShut
  {NoStop}%
\bibitem [{\citenamefont {Anselmino}\ \emph {et~al.}(2015)\citenamefont
  {Anselmino} \emph {et~al.}}]{Anselmino:2015sxa}%
  \BibitemOpen
  \bibfield  {author} {\bibinfo {author} {\bibfnamefont {M.}~\bibnamefont
  {Anselmino}} \emph {et~al.},\ }\href
  {https://doi.org/10.1103/PhysRevD.92.114023} {\bibfield  {journal} {\bibinfo
  {journal} {Phys. Rev. D}\ }\textbf {\bibinfo {volume} {92}},\ \bibinfo
  {pages} {114023} (\bibinfo {year} {2015})},\ \Eprint
  {https://arxiv.org/abs/1510.05389} {arXiv:1510.05389 [hep-ph]} \BibitemShut
  {NoStop}%
\bibitem [{\citenamefont {Ethier}\ \emph {et~al.}(2017)\citenamefont {Ethier},
  \citenamefont {Sato},\ and\ \citenamefont {Melnitchouk}}]{Ethier:2017zbq}%
  \BibitemOpen
  \bibfield  {author} {\bibinfo {author} {\bibfnamefont {J.~J.}\ \bibnamefont
  {Ethier}}, \bibinfo {author} {\bibfnamefont {N.}~\bibnamefont {Sato}},\ and\
  \bibinfo {author} {\bibfnamefont {W.}~\bibnamefont {Melnitchouk}},\ }\href
  {https://doi.org/10.1103/PhysRevLett.119.132001} {\bibfield  {journal}
  {\bibinfo  {journal} {Phys. Rev. Lett.}\ }\textbf {\bibinfo {volume} {119}},\
  \bibinfo {pages} {132001} (\bibinfo {year} {2017})},\ \Eprint
  {https://arxiv.org/abs/1705.05889} {arXiv:1705.05889 [hep-ph]} \BibitemShut
  {NoStop}%
\bibitem [{\citenamefont {Martin}\ \emph {et~al.}(2017)\citenamefont {Martin},
  \citenamefont {Bradamante},\ and\ \citenamefont {Barone}}]{Martin:2017yms}%
  \BibitemOpen
  \bibfield  {author} {\bibinfo {author} {\bibfnamefont {A.}~\bibnamefont
  {Martin}}, \bibinfo {author} {\bibfnamefont {F.}~\bibnamefont {Bradamante}},\
  and\ \bibinfo {author} {\bibfnamefont {V.}~\bibnamefont {Barone}},\ }\href
  {https://doi.org/10.1103/PhysRevD.95.094024} {\bibfield  {journal} {\bibinfo
  {journal} {Phys. Rev. D}\ }\textbf {\bibinfo {volume} {95}},\ \bibinfo
  {pages} {094024} (\bibinfo {year} {2017})},\ \Eprint
  {https://arxiv.org/abs/1701.08283} {arXiv:1701.08283 [hep-ph]} \BibitemShut
  {NoStop}%
\bibitem [{\citenamefont {Abbon}\ \emph {et~al.}(2007)\citenamefont {Abbon}
  \emph {et~al.}}]{Abbon:2007pq}%
  \BibitemOpen
  \bibfield  {author} {\bibinfo {author} {\bibfnamefont {P.}~\bibnamefont
  {Abbon}} \emph {et~al.} (\bibinfo {collaboration} {COMPASS collaboration}),\
  }\href {https://doi.org/10.1016/j.nima.2007.03.026} {\bibfield  {journal}
  {\bibinfo  {journal} {Nucl. Instrum. Meth. A}\ }\textbf {\bibinfo {volume}
  {577}},\ \bibinfo {pages} {455} (\bibinfo {year} {2007})},\ \Eprint
  {https://arxiv.org/abs/hep-ex/0703049} {arXiv:hep-ex/0703049} \BibitemShut
  {NoStop}%
\bibitem [{\citenamefont {Adolph}\ \emph {et~al.}(2015)\citenamefont {Adolph}
  \emph {et~al.}}]{Adolph:2014zba}%
  \BibitemOpen
  \bibfield  {author} {\bibinfo {author} {\bibfnamefont {C.}~\bibnamefont
  {Adolph}} \emph {et~al.} (\bibinfo {collaboration} {COMPASS collaboration}),\
  }\href {https://doi.org/10.1016/j.physletb.2015.03.056} {\bibfield  {journal}
  {\bibinfo  {journal} {Phys. Lett. B}\ }\textbf {\bibinfo {volume} {744}},\
  \bibinfo {pages} {250} (\bibinfo {year} {2015})},\ \Eprint
  {https://arxiv.org/abs/1408.4405} {arXiv:1408.4405 [hep-ex]} \BibitemShut
  {NoStop}%
\bibitem [{\citenamefont {Ageev}\ \emph {et~al.}(2005)\citenamefont {Ageev}
  \emph {et~al.}}]{COMPASS:2005xxc}%
  \BibitemOpen
  \bibfield  {author} {\bibinfo {author} {\bibfnamefont {E.~S.}\ \bibnamefont
  {Ageev}} \emph {et~al.} (\bibinfo {collaboration} {COMPASS collaboration}),\
  }\href {https://doi.org/10.1016/j.physletb.2005.03.025} {\bibfield  {journal}
  {\bibinfo  {journal} {Phys. Lett. B}\ }\textbf {\bibinfo {volume} {612}},\
  \bibinfo {pages} {154} (\bibinfo {year} {2005})},\ \Eprint
  {https://arxiv.org/abs/hep-ex/0501073} {arXiv:hep-ex/0501073} \BibitemShut
  {NoStop}%
\bibitem [{\citenamefont {Maguire}\ \emph {et~al.}(2017)\citenamefont
  {Maguire}, \citenamefont {Heinrich},\ and\ \citenamefont
  {Watt}}]{Maguire:2017ypu}%
  \BibitemOpen
  \bibfield  {author} {\bibinfo {author} {\bibfnamefont {E.}~\bibnamefont
  {Maguire}}, \bibinfo {author} {\bibfnamefont {L.}~\bibnamefont {Heinrich}},\
  and\ \bibinfo {author} {\bibfnamefont {G.}~\bibnamefont {Watt}},\ }\href
  {https://doi.org/10.1088/1742-6596/898/10/102006} {\bibfield  {journal}
  {\bibinfo  {journal} {J. Phys. Conf. Ser.}\ }\textbf {\bibinfo {volume}
  {898}},\ \bibinfo {pages} {102006} (\bibinfo {year} {2017})},\ \Eprint
  {https://arxiv.org/abs/1704.05473} {arXiv:1704.05473 [hep-ex]} \BibitemShut
  {NoStop}%
\bibitem [{\citenamefont {Augsten}\ \emph {et~al.}(2018)\citenamefont {Augsten}
  \emph {et~al.}}]{COMPASS:2018shj}%
  \BibitemOpen
  \bibfield  {author} {\bibinfo {author} {\bibfnamefont {K.}~\bibnamefont
  {Augsten}} \emph {et~al.} (\bibinfo {collaboration} {COMPASS
  collaboration}),\ }\href@noop {} {\bibfield  {journal} {\bibinfo  {journal}
  {SPSC-P-340-ADD-1, CERN–SPSC–2017–034}\ } (\bibinfo {year}
  {2018})}\BibitemShut {NoStop}%
\end{thebibliography}
%

\clearpage

\center{\textbf{The COMPASS Collaboration}}

\vspace{10pt}
\begin{flushleft}
G.~D.~Alexeev$^\textrm{{\footnotesize\hyperlink{hl:dubna}{29}}}$\orcidlink{0009-0007-0196-8178},
M.~G.~Alexeev$^\textrm{{\footnotesize\hyperlink{hl:turin_u}{20},\hyperlink{hl:turin_i}{19}}}$\orcidlink{0000-0002-7306-8255},
C.~Alice$^\textrm{{\footnotesize\hyperlink{hl:turin_u}{20},\hyperlink{hl:turin_i}{19}}}$\orcidlink{0000-0001-6297-9857},
A.~Amoroso$^\textrm{{\footnotesize\hyperlink{hl:turin_u}{20},\hyperlink{hl:turin_i}{19}}}$\orcidlink{0000-0002-3095-8610},
V.~Andrieux$^\textrm{{\footnotesize\hyperlink{hl:illinois}{34}}}$\orcidlink{0000-0001-9957-9910},
V.~Anosov$^\textrm{{\footnotesize\hyperlink{hl:dubna}{29}}}$\orcidlink{0009-0003-3595-9561},
S.~Asatryan$^\textrm{{\footnotesize\hyperlink{hl:aanl}{1}}}$\orcidlink{0000-0002-4325-3329},
K.~Augsten$^\textrm{{\footnotesize\hyperlink{hl:praguectu}{4}}}$\orcidlink{0000-0001-8324-0576},
W.~Augustyniak$^\textrm{{\footnotesize\hyperlink{hl:warsaw}{24}}}$,
C.~D.~R.~Azevedo$^\textrm{{\footnotesize\hyperlink{hl:aveiro}{27}}}$\orcidlink{0000-0002-0012-9918}
% ...
\hypertarget{hl:calcutta}{$^\textrm{{\footnotesize 14}}$\footnotesize~Matrivani Institute of Experimental Research \& Education, Calcutta-700 030, India$^\textrm{{\tiny\hyperlink{hl:C}{C}}}$\\}
\hypertarget{hl:telaviv}{$^\textrm{{\footnotesize 15}}$\footnotesize~Tel Aviv University, School of Physics and Astronomy, 69978 Tel Aviv, Israel$^\textrm{{\tiny\hyperlink{hl:D}{D}}}$\\}
\hypertarget{hl:triest_a}{$^\textrm{{\footnotesize 16}}$\footnotesize~Abdus Salam ICTP, 34151 Trieste, Italy\\}
\hypertarget{hl:triest_i}{$^\textrm{{\footnotesize 17}}$\footnotesize~Trieste Section of INFN, 34127 Trieste, Italy\\}
\hypertarget{hl:triest_u}{$^\textrm{{\footnotesize 18}}$\footnotesize~University of Trieste, Dept.\ of Physics, 34127 Trieste, Italy\\}
\hypertarget{hl:turin_i}{$^\textrm{{\footnotesize 19}}$\footnotesize~Torino Section of INFN, 10125 Torino, Italy\\}
\hypertarget{hl:turin_u}{$^\textrm{{\footnotesize 20}}$\footnotesize~University of Torino, Dept.\ of Physics, 10125 Torino, Italy\\}
\hypertarget{hl:miyazaki}{$^\textrm{{\footnotesize 21}}$\footnotesize~University of Miyazaki, Miyazaki 889-2192, Japan$^\textrm{{\tiny\hyperlink{hl:E}{E}}}$\\}
\hypertarget{hl:nagoya}{$^\textrm{{\footnotesize 22}}$\footnotesize~Nagoya University, 464 Nagoya, Japan$^\textrm{{\tiny\hyperlink{hl:E}{E}}}$\\}
\hypertarget{hl:yamagata}{$^\textrm{{\footnotesize 23}}$\footnotesize~Yamagata University, Yamagata 992-8510, Japan$^\textrm{{\tiny\hyperlink{hl:E}{E}}}$\\}
\hypertarget{hl:warsaw}{$^\textrm{{\footnotesize 24}}$\footnotesize~National Centre for Nuclear Research, 02-093 Warsaw, Poland$^\textrm{{\tiny\hyperlink{hl:F}{F}}}$\\}
\hypertarget{hl:warsawtu}{$^\textrm{{\footnotesize 25}}$\footnotesize~Warsaw University of Technology, Institute of Radioelectronics, 00-665 Warsaw, Poland$^\textrm{{\tiny\hyperlink{hl:F}{F}}}$\\}
\hypertarget{hl:warsawu}{$^\textrm{{\footnotesize 26}}$\footnotesize~University of Warsaw, Faculty of Physics, 02-093 Warsaw, Poland$^\textrm{{\tiny\hyperlink{hl:F}{F}}}$\\}
\hypertarget{hl:aveiro}{$^\textrm{{\footnotesize 27}}$\footnotesize~University of Aveiro, I3N, Dept. of Physics, 3810-193 Aveiro, Portugal$^\textrm{{\tiny\hyperlink{hl:G}{G}}}$\\}
\hypertarget{hl:lisbon}{$^\textrm{{\footnotesize 28}}$\footnotesize~LIP, 1649-003 Lisbon, Portugal$^\textrm{{\tiny\hyperlink{hl:G}{G}}}$\\}
\hypertarget{hl:dubna}{$^\textrm{{\footnotesize 29}}$\footnotesize~Affiliated with an international laboratory covered by a cooperation agreement with CERN\\}
\hypertarget{hl:russia}{$^\textrm{{\footnotesize 30}}$\footnotesize~Affiliated with an institute covered by a cooperation agreement with CERN.\\}
\hypertarget{hl:cern}{$^\textrm{{\footnotesize 31}}$\footnotesize~CERN, 1211 Geneva 23, Switzerland\\}
\hypertarget{hl:taipei}{$^\textrm{{\footnotesize 32}}$\footnotesize~Academia Sinica, Institute of Physics, Taipei 11529, Taiwan$^\textrm{{\tiny\hyperlink{hl:H}{H}}}$\\}
\hypertarget{hl:taipeincu}{$^\textrm{{\footnotesize 33}}$\footnotesize~Center for High Energy and High Field Physics and Dept.\ of Physics, National Central University, 300 Zhongda Rd., Zhongli 320317, Taiwan$^\textrm{{\tiny\hyperlink{hl:H}{H}}}$\\}
\hypertarget{hl:illinois}{$^\textrm{{\footnotesize 34}}$\footnotesize~University of Illinois at Urbana-Champaign, Dept.\ of Physics, Urbana, IL 61801-3080, USA$^\textrm{{\tiny\hyperlink{hl:I}{I}}}$\\}

\vspace{10pt}
\hypertarget{hl:*}{$^\textrm{{\footnotesize *}}$\footnotesize~Corresponding author\\}
\hypertarget{hl:a}{$^\textrm{{\footnotesize a}}$\footnotesize~Supported by the European Union’s Horizon 2020 research and innovation programme under grant agreement STRONG–2020 - No 824093\\}
\hypertarget{hl:b}{$^\textrm{{\footnotesize b}}$\footnotesize~Retired from Ludwig-Maximilians-Universit\"at, 80539 M\"unchen, Germany\\}
\hypertarget{hl:b1}{$^\textrm{{\footnotesize b1}}$\footnotesize~Supported by the DFG cluster of excellence `Origin and Structure of the Universe' (www.universe-cluster.de) (Germany)\\}
\hypertarget{hl:c}{$^\textrm{{\footnotesize c}}$\footnotesize~Also at ORIGINS Excellence Cluster, 85748 Garching, Germany\\}
\hypertarget{hl:d}{$^\textrm{{\footnotesize d}}$\footnotesize~Also at Institut f\"ur Theoretische Physik, Universit\"at T\"ubingen, 72076 T\"ubingen, Germany\\}
\hypertarget{hl:e}{$^\textrm{{\footnotesize e}}$\footnotesize~Present address: NISER, Centre for Medical and Radiation Physics, Bubaneswar, India\\}
\hypertarget{hl:f}{$^\textrm{{\footnotesize f}}$\footnotesize~Also at University of Eastern Piedmont, 15100 Alessandria, Italy\\}
\hypertarget{hl:f1}{$^\textrm{{\footnotesize f1}}$\footnotesize~Supported by the Funds for Research 2019-22 of the University of Eastern Piedmont\\}
\hypertarget{hl:g}{$^\textrm{{\footnotesize g}}$\footnotesize~Also at Chubu University, Kasugai, Aichi 487-8501, Japan\\}
\hypertarget{hl:h}{$^\textrm{{\footnotesize h}}$\footnotesize~Also at KEK, 1-1 Oho, Tsukuba, Ibaraki 305-0801, Japan\\}
\hypertarget{hl:i}{$^\textrm{{\footnotesize i}}$\footnotesize~Also at Dept.\ of Physics, Pusan National University, Busan 609-735, Republic of Korea\\}
\hypertarget{hl:i1}{$^\textrm{{\footnotesize i1}}$\footnotesize~Also at Physics Dept., Brookhaven National Laboratory, Upton, NY 11973, USA\\}
\hypertarget{hl:j}{$^\textrm{{\footnotesize j}}$\footnotesize~Also at Fairmont State University, Department of Natural Sciences, 1201 Locust Ave, Fairmont, West Virginia 26554, USA\\}
\hypertarget{hl:k}{$^\textrm{{\footnotesize k}}$\footnotesize~Also at Dept.\ of Physics, National Kaohsiung Normal University, Kaohsiung County 824, Taiwan\\}
\hypertarget{hl:$\dagger$}{$^\textrm{{\footnotesize $\dagger$}}$\footnotesize~Deceased\\}

\vspace{10pt}
\hypertarget{hl:A}{$^\textrm{{\tiny A}}$\footnotesize~Supported by MEYS Grants LM2018104, LM2023040 and LTT17018 and Charles University grants PRIMUS/22/SCI/017 and GAUK60121 (Czech Republic)\\}
\hypertarget{hl:B}{$^\textrm{{\tiny B}}$\footnotesize~Supported by BMBF - Bundesministerium f\"ur Bildung und Forschung (Germany)\\}
\hypertarget{hl:C}{$^\textrm{{\tiny C}}$\footnotesize~Supported by B. Sen fund (India)\\}
\hypertarget{hl:D}{$^\textrm{{\tiny D}}$\footnotesize~Supported by the Israel Academy of Sciences and Humanities (Israel)\\}
\hypertarget{hl:E}{$^\textrm{{\tiny E}}$\footnotesize~Supported by MEXT and JSPS, Grants 18002006, 20540299, 18540281 and 26247032, the Daiko and Yamada Foundations (Japan)\\}
\hypertarget{hl:F}{$^\textrm{{\tiny F}}$\footnotesize~Supported by NCN, Grant 2020/37/B/ST2/01547 (Poland)\\}
\hypertarget{hl:G}{$^\textrm{{\tiny G}}$\footnotesize~Supported by FCT, Grants DOI 10.54499/CERN/FIS-PAR/0022/2019 and DOI 10.54499/CERN/FIS-PAR/0016/2021 (Portugal)\\}
\hypertarget{hl:H}{$^\textrm{{\tiny H}}$\footnotesize~Supported by the Ministry of Science and Technology (Taiwan)\\}
\hypertarget{hl:I}{$^\textrm{{\tiny I}}$\footnotesize~Supported by the National Science Foundation, Grant no. PHY-1506416 (USA)\\}

\end{flushleft}

\end{document}